\documentclass[pdftex,twocolumn,epjc3]{svjour3}          

\RequirePackage[T1]{fontenc}

\smartqed  

\RequirePackage{graphicx}
\RequirePackage{mathptmx}      
\RequirePackage{flushend}
\RequirePackage[numbers,sort&compress]{natbib}
\RequirePackage[colorlinks,citecolor=blue,urlcolor=blue,linkcolor=blue]{hyperref}

\journalname{Eur. Phys. J. A}

\begin{document}

\title{An innovative technique for the investigation of the 4-fold forbidden beta-decay of $^{50}$V}


\author{L.~Pattavina\thanksref{GSSI,LNGS,e1}
        \and
       M.~Laubenstein\thanksref{LNGS}\and
      S.S.~Nagorny\thanksref{LNGS,e2} \and
       S.~Nisi\thanksref{LNGS} \and
       L.~Pagnanini\thanksref{LNGS} \and
      S.~Pirro\thanksref{LNGS} \and
      C.~Rusconi\thanksref{LNGS,UCS} \and
      K.~Sch\"affner\thanksref{GSSI,LNGS} 
}

\thankstext{e1}{Corresponding author: luca.pattavina@lngs.infn.it}
\thankstext{e2}{Present address: Queen's University, K7L 3N6 Kingston, Canada}

\institute{Gran Sasso Science Institute, L'Aquila I-67100 - Italy\label{GSSI} 
\and
INFN - Laboratori Nazionali del Gran Sasso, Assergi (L'Aquila) I-67100 - Italy\label{LNGS}
          \and
         Dept. of Physics and Astronomy, University of South Carolina, SC 29208, Columbia, USA\label{UCS}
}

\date{\today}

\maketitle

\begin{abstract}
For the first time a Vanadium-based crystal was operated as cryogenic particle detector. The scintillating low temperature calorimetric technique was used for the characterization of a 22~g YVO$_4$ crystal aiming at the investigation of the 4-fold forbidden non-unique $\beta^-$ decay of $^{50}$V. The excellent bolometric performance of the compound together with high light output of the crystal makes it an outstanding technique for the study of such elusive rate process. The internal radioactive contaminations of the crystal are also investigated showing that an improvement on the current status of material selection and purification are needed, $^{235/238}$U and $^{232}$Th are measured at the level of 28~mBq/kg, 1.3~Bq/kg and 28~mBq/kg, respectively. In this work, we also discuss a future upgrade of the experimental set-up which may pave the road for the detection of the rare $^{50}$V $\beta^-$ decay.
\end{abstract}

\keywords{Beta decays, cryogenic detectors, particle discrimination}

\section{Introduction}
\label{intro}

In nature there are a number of isotopes, which are considered as {\it stable}, nevertheless their radioactive decay is energetically allowed. Among these a valuable example is given by $^{209}$Bi, which was supposed to be stable until 2003, when its $\alpha$ decay~\cite{Bi209,Bi209exc} was observed for the first time. However, their rate of decay is largely suppressed by selection rules for angular momentum and parity between the initial and the final states of the nucleus.

The naturally occurring Vanadium isotope $^{50}$V is one of three isotopes, together with $^{113}$Cd and $^{115}$In, for which a non-unique 4-fold forbidden $\beta^-$-decay may occur - parity of the decay does not change, while the angular momentum changes by four units $\Delta l ^{\Delta\pi}$= 4$^+$.

The study of $^{50}$V decay is a useful probe to clarify the crucial problem of the quenching of the weak axial-vector
coupling constant g$_{A}$ in the nuclear media~\cite{iachello}. A study of the spectral shape of the decay may shade light on the value of g$_A$, as suggested in~\cite{Suhonen}.

\begin{figure}[b]
\centering
\includegraphics[width=0.4\textwidth]{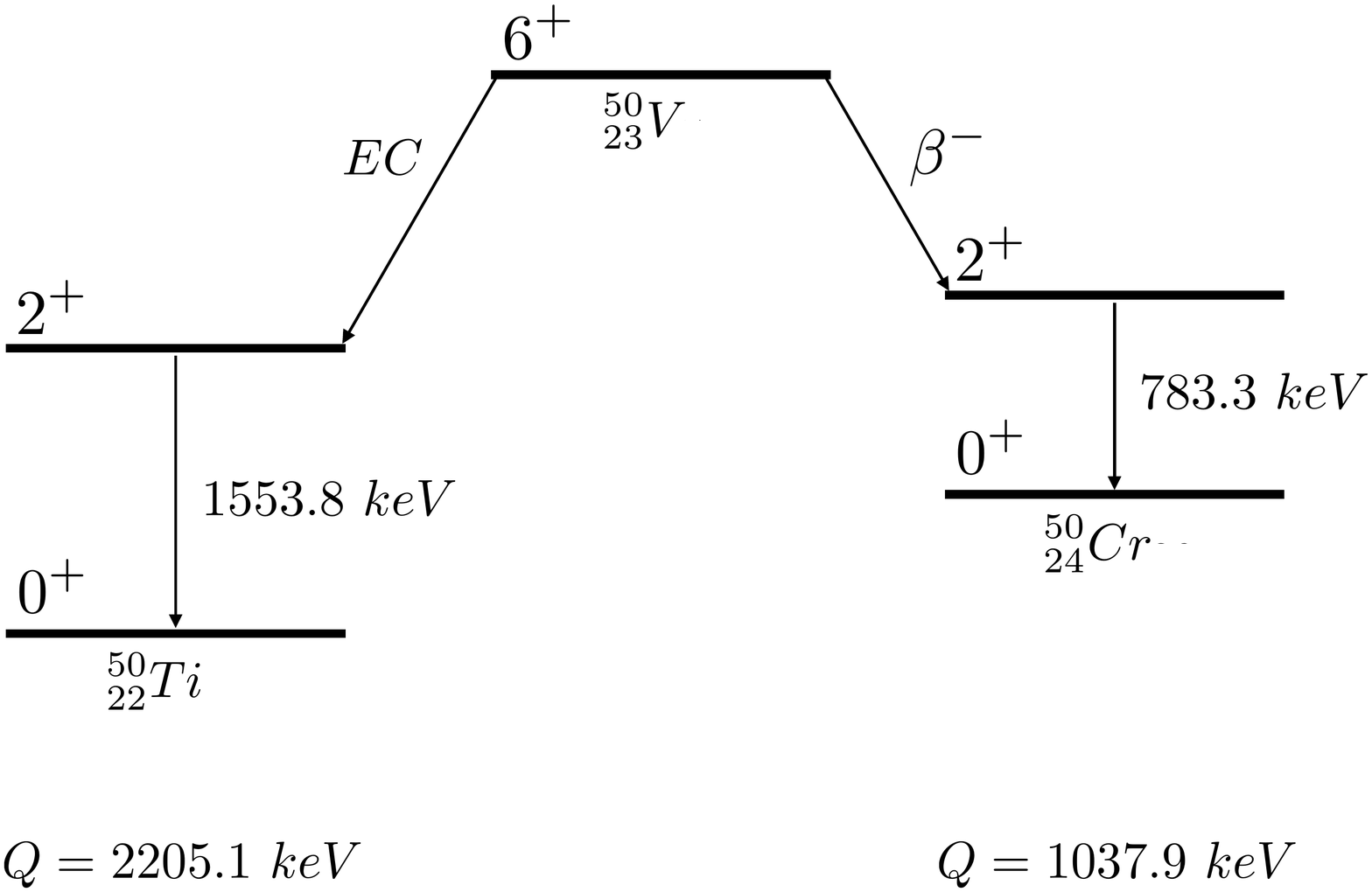}
\caption{$^{50}$V decay scheme.}
\label{fig:scheme}
\end{figure}

In the last 60 years many studies have been carried out for the investigation of $^{50}$V decay \cite{1V,2V,3V,4V,5V,6V,7V,8V,9V,10V,11V}. The isotope is very interesting, since the decay to the ground state of the daughter nucleus is strongly suppressed, being a 6-fold forbidden decay. Thus, $^{50}$V can either undergo 4-fold forbidden $\beta^-$ decay to the first excited state of $^{50}$Cr or it can decay through electron capture to the first excited level of $^{50}$Ti, see Fig.~\ref{fig:scheme}. 

The estimated theoretical half-life of this nucleus is $>$10$^{17}$~y~\cite{1V}. For a detector of few kilograms in zero-background conditions, only few events per month are expected. This makes the investigation of this rare process extremely difficult, given the elusive counting rate. In 1989, J.J.~Simpson et al.~\cite{10V} measured for the first time the half-life of $^{50}$V electron capture decay to be (2.05$\pm$0.49)$\times$10$^{17}$~y, while up to now no clear observation of the $^{50}$V $\beta^-$ decay has been found. The strongest limit on the half-life of this decay is set at $>$8$\times$10$^{17}$~y~\cite{11V}.

\section{Scintillating bolometers for rare event investigations}
\label{sec:1}
Significant progress has been made in the field of rare decay investigation during the past decade exploiting the scintillating bolometric technique~\cite{Pirro_rev}. After the discovery of the $^{209}$Bi $\alpha$ decay (T$_{1/2}$ = 1.9$\times$10$^{19}$~y) in 2003~\cite{Bi209}, the $\alpha$ decay of $^{180}$W~\cite{W180} to the ground state and the decay of $^{209}$Bi to the excited state~\cite{Bi209exc} were detected few years later. Then, the most stringent experimental limits on the half-lives of Pb isotopes $\alpha$ decay were set~\cite{pbwo4}. Recently $^{151}$Eu and $^{148}$Sm $\alpha$ decays were observed~\cite{Eu151,Sm148}. The main benefit for using scintillating bolometers  for the investigation of rare $\alpha$ decays is to have the detector absorber made of the decay source. In all the previously mentioned measurements the scintillating bolometer detector was made of the decay source, thus ensuring high detection efficiency at the level of $>$95\%. In this work, we used an undoped YVO$_4$ single crystal as bolometer for the feasibility of studying of $^{50}$V $\beta^-$ decay.

We discuss the crystal properties in terms of light yield and internal radiopurity, as well as its bolometric performance in terms of energy resolution and signal amplitude.

\subsection{Crystal characteristics}
YVO$_4$ is an attractive compound given its large mass fraction of Vanadium (about 18\%). The crystal production technology is well-established, conventional Czochralski method, and it is available on the market. Nevertheless, only small size crystal are purchasable with volume around 7~cm$^3$. The crystal has good mechanical properties and it features a wide transparency range from 400~nm up to 3~$\mu$m, for these reasons it is largely employed in laser applications.

The crystal characterized in this work was purchased on the market without no specifications on its properties, other than being undoped. It weighs 22~g and it has a cylindrical shape of 18~mm of diameter and 20~mm of height.

\subsection{Detector working principle}
A cryogenic particle detector, namely a bolometer, is a crystal absorber made of a proper material, which is directly coupled to a suitable thermometer. If a particle traverses the absorber and releases energy, a temperature rise is induced in the crystal and measured by the thermometer. In order to reduce the temperature fluctuations and to enhance the signal amplitude, the experimental set-up is operated at cryogenic temperature so that the heat capacity of the absorber is reduced to the level where a single particle interaction can induce a sizeable positive temperature variation.

When the crystal absorber is also an efficient scintillator at low temperature, then a fraction of the deposited energy is transformed into a scintillation signal which can be measured by means of a suitable light detector (LD). The double heat-light read-out allows for an efficient identification of the nature of the interacting particle, using the Light Yield (LY) discrimination~\footnote{We define the Light Yield as the ratio between the measured light, in keV, and the nominal energy of an event, in MeV.}.

Scintillating bolometers are outstanding devices for the investigation of rare nuclear processes such as the 4-fold forbidden $\beta^-$ decay of $^{50}$V. In fact, they ensures high detection efficiency - the detector is made of the decay source - low background - using the double read-out heat and light - and excellent energy resolution over a wide energy range - at the level of permille.

\subsection{Experimental set-up}

\begin{figure}[h!]
\centering
\includegraphics[width=0.4\textwidth]{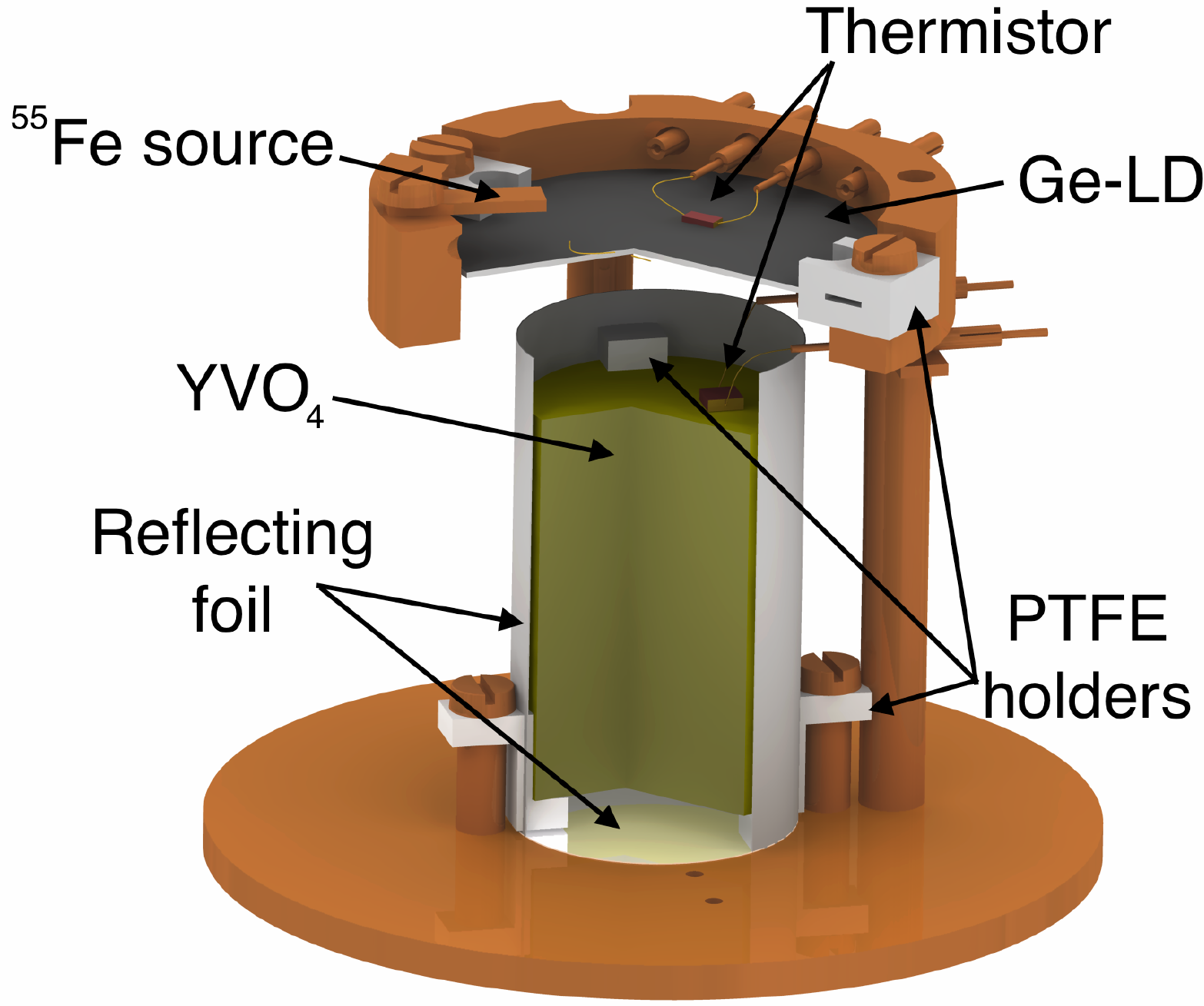}
\caption{Detector's set-up. The YVO$_4$ crystal is surrounded by VIKUITI-3M reflecting foil, except for its top side which is faced by a cryogenic Ge Light Detector. Light and heat signals are both read by means of two thermistors coupled on the absorbers. The whole structure is housed in a high purity copper structure. A permanent $^{55}$Fe source is mounted facing the LD for its energy calibration.}
\label{fig:setup}
\end{figure}

A 22~g crystal of YVO$_4$ was enclosed in a high-purity Cu structure, following the same procedure adopted in~\cite{LiMoNe}, where the crystal is kept in position by means of PTFE clamps, which act also as heat sink for keeping the crystal at cryogenic temperature, as shown in Fig.~\ref{fig:setup}. The YVO$_4$ absorber is fully surrounded, except for the top face, by a VIKUITI-3M$^{\textregistered}$ reflecting foil, which guarantees an efficient light collection to the LD. 

The LD employed for the detection of the scintillation light produced in the YVO$_4$ crystal is a thin wafer of High-Purity Germanium operated as auxiliary bolometer in coincidence with the main crystal. More details on operation of LD's can be found in~\cite{LD_characterization}. Ge-NTD (Neutron Transmutation Doped) thermistors are used as thermal sensors for the measurement of the temperature rise induced by particle interactions in the absorbers. The signals are read from the voltage drop across the electrical connections on the thermistors.

The YVO$_4$ and its LD were installed in a $^3$He/$^4$He dilution refrigerator at the underground Laboratori del Gran Sasso of INFN. This location guarantees a muon flux reduction of about a factor 10$^6$ compared to sea level. Reducing the particle flux that can interact with the experimental set-up is mandatory while operating bolometers due to their slow signal development of few seconds~\cite{muon}.

To further reduce the rate of interactions in the absorber, the cryostat is shielded against natural radioactivity with 10~cm of low-radioactivity Pb (50~Bq/kg) and a multilayer of paraffin and borated polyethylene for the thermalization and absorption of high energy neutrons. The whole set-up is enclosed inside an anti-Radon box continuously flushed with boil-off N$_2$ gas.

During the data taking the acquired signals are amplified by JFETs and then fed into a 6 poles low-pass Bessel filter, and finally recorded by an 18~bit NI-6284 PXI ADC unit. For more details on the electronics contribution on the detector performance, see~\cite{Zn82Se}.

The YVO$_4$ absorber and the LD waveforms are sampled at 1~kHz frequency. The software trigger is set so that 250~ms and 125~ms long time-windows are acquired when it fires on the YVO$_4$ and on the LD respectively. Moreover when the trigger on the YVO$_4$ is fired, the corresponding waveform from the LD is always recorded, irrespectively of its trigger. This allows for an efficient synchronization of the heat and light signals~\cite{light_sync}.

\begin{figure}[b]
\centering
\includegraphics[width=0.5\textwidth]{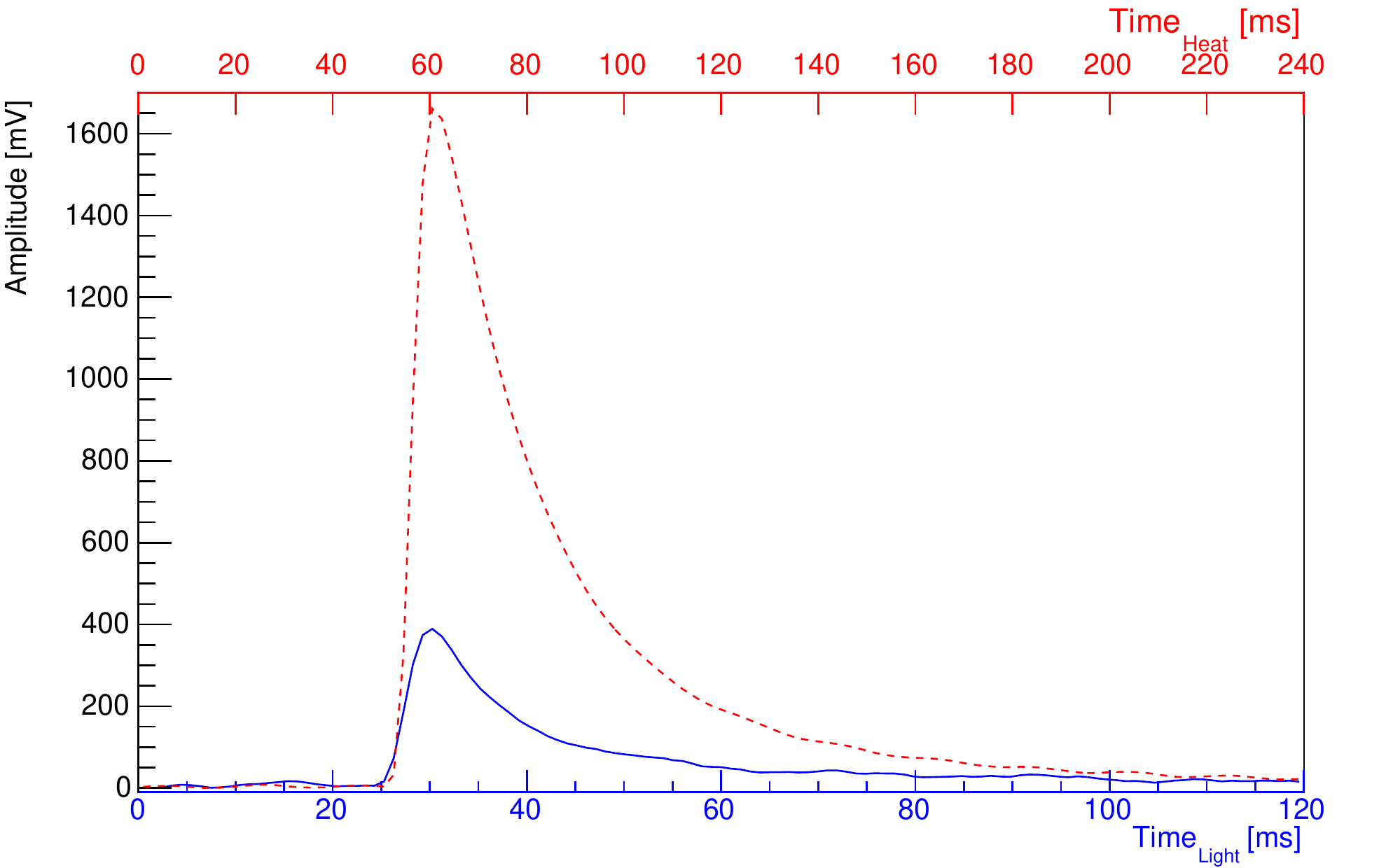}
\caption{Synchronization of heat and light signals induced by a $\beta$/$\gamma$ event. The pulse with the dashed line is acquired on the YVO$_4$ detector, while the one with a solid line is the corresponding scintillation light signal detected on the LD. The two pulses have different acquisition time-windows.}
\label{fig:pulso}
\end{figure}

In Fig.~\ref{fig:pulso} is shown a standard $\beta$/$\gamma$ event on the YVO$_4$ crystal and its corresponding light signal acquired on the LD. The LD signal develops few ms before the one of the YVO$_4$, given the different heat capacities of the two systems, smaller for the LD and larger for the YVO$_4$. 

\section{Data Processing and Analysis}

A first level of analysis is performed on the acquired data, over more than 120~h of background and calibration runs. This consists in filtering the triggered pulses by means of the Optimum Filter technique~\cite{OF}, which maximizes the signal-to-noise ratio analysing the characteristic frequency of the acquired pulses and of the noise. The role of the filter is to weigh the frequency components of the signal in order to suppress those frequencies that are more affected by noise. Once the optimal signal amplitude is estimated, the gain instabilities of the detector are corrected. This procedure is carried out exploiting the fixed energy reference thermal pulses produced on the detector by means of the heater resistor coupled to the absorbers. The last operation performed in this first-level analysis is the conversion of the stabilized optimal amplitudes into the corresponding energy values. During the data taking, specially designed $^{232}$Th sources were inserted in the cryogenic system for calibrating the detector energy response. The LD, due to their small mass, is calibrated by means of permanent X-rays $^{55}$Fe source facing the Ge wafer.

The energy spectra for the two detectors are composed by events which passed two classes of data selection. The first one is meant to exclude excess of noise from the data analysis (e.g. external electromagnetic interferences or irreducible mechanical vibrations of the system). The second cut selects only particle-induced events rejecting non-physical triggered data (e.g. electronic spikes). This selection is performed through a pulse-shape analysis, comparing template pulses with the acquired ones.

In Table~\ref{tab:eff} are reported the efficiencies for the pulse-shape data selection cuts and the pile-up rejection efficiency. The low efficiency on the pulse shape cuts is mainly due to the high count rate of the detector which is evaluated to be about 120~mHz. The efficiencies are computed following the same procedure adopted in~\cite{pbwo4}.

\begin{table}
\begin{center}
\caption{Data selection efficiencies for the main $\alpha$ peaks: pulse-shape cut, pile-up rejection and total efficiency.} 
\begin{tabular}{lcccc}
\hline\noalign{\smallskip}
Nuclide & Q-value & $\epsilon_{PS}$ &  $\epsilon_{Pile-up}$ & $\epsilon_{TOT}$ \\ 
\noalign{\smallskip}\hline\noalign{\smallskip}
$^{232}$Th & 4082~keV & 0.44$\pm$0.01  & 0.97$\pm$0.01 & 0.43$\pm$0.02 \\
$^{238}$U & 4270~keV & 0.45$\pm$0.01  & 0.97$\pm$0.01 & 0.44$\pm$0.02 \\
$^{234}$U & 4858~keV & 0.46$\pm$0.01  & 0.97$\pm$0.01 & 0.45$\pm$0.02 \\
$^{226}$Ra & 4871~keV & 0.40$\pm$0.02  & 0.97$\pm$0.01 & 0.39$\pm$0.02 \\
$^{210}$Po & 5407~keV & 0.43$\pm$0.02  & 0.97$\pm$0.01 & 0.42$\pm$0.02 \\
$^{235}$U & 4678~keV & 0.33$\pm$0.02  & 0.97$\pm$0.01 & 0.32$\pm$0.02 \\
$^{147}$Sm & 2516~keV & 0.47$\pm$0.02 & 0.97$\pm$0.01 & 0.46$\pm$0.02 \\
\noalign{\smallskip}\hline
\end{tabular}
\label{tab:eff}
\end{center}
\end{table}

\section{Detector perfomance}

The detector bolometric performance parameters are reported in Tab.~\ref{tab:bolo}. The FWHM$_{0keV}$ represents the noise fluctuation at the filtered output, which is mainly caused by microphonic noise of the cryogenic infrastructure. It does not depend on the absolute signal amplitude. The rise ($\tau_{rise}$) and decay ($\tau_{decay}$) times are computed as the time interval between the 10\% and 90\% of the leading edge of the pulse amplitude and as the 90\% and 30\% of the trailing part of the pulse amplitude, respectively.

The large signal amplitude and the fast time response of the crystal makes it a good candidate for cryogenic operation.

\begin{table}
\begin{center}
\caption{Bolometric performance of the detector set-up: YVO$_4$ and Light Detector. The signal amplitude refers to the sensitivity of the detectors, the FWHM$_{0keV}$ to the baseline energy resolution. $\tau_{rise/decay}$ are the rise time and decay time computed as the time interval between the 10\% and 90\% of the leading edge of the pulse amplitude and as the 90\% and 30\% of the trailing part of the pulse amplitude, respectively. } 
\begin{tabular}{lcccc}
\hline\noalign{\smallskip}
Crystal & Signal Amplitude & FWHM$_{0keV}$ &  $\tau_{rise}$ & $\tau_{decay}$ \\ 
            &  [$\mu$V/MeV]& [keV] &  [ms] & [ms]\\
\noalign{\smallskip}\hline\noalign{\smallskip}
YVO$_4$ & 215 & 1.22 & 6 & 25    \\
LD & 770 & 0.19 & 3 & 10\\
\noalign{\smallskip}\hline
\end{tabular}
\label{tab:bolo}
\end{center}
\end{table}

\begin{figure}[h]
\centering
\includegraphics[width=0.5\textwidth]{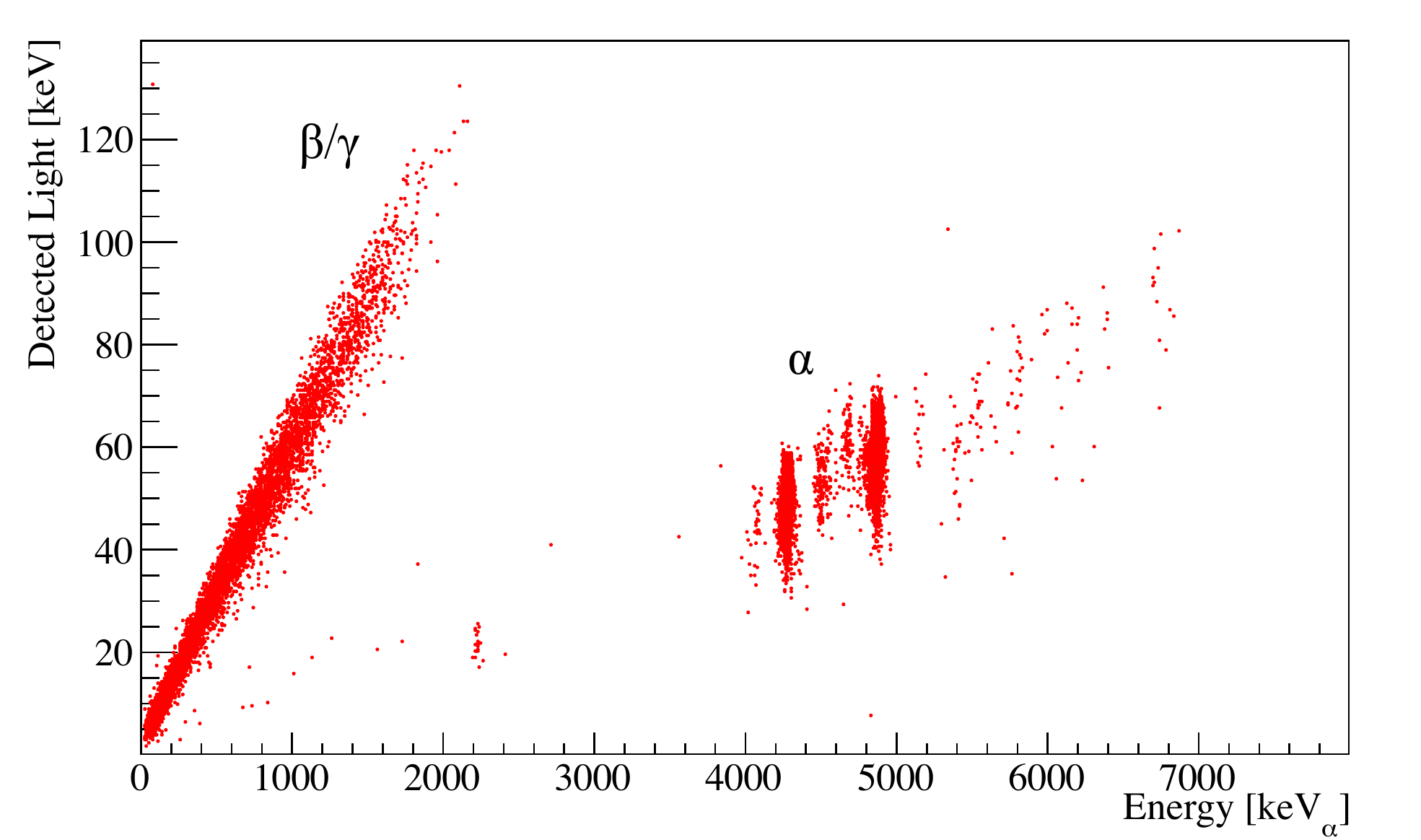}
\caption{Energy scatter plot Heat vs. Light. The X-axis shows the energy released in the YVO$_4$ absorber and it is energy-calibrated using the internal $\alpha$ radioactive contamination, while the Y-axis shows the light signal induced in the Light Detector.}
\label{fig:scatter}
\end{figure}

The data acquired during the entire 115~h of the background run are shown in Fig.~\ref{fig:scatter}. Each point in the plot represents an event occurring in the YVO$_4$ crystal and the corresponding induced light signal in the LD. We can identify clearly two bands, one produced by $\beta$/$\gamma$ interaction which extends from the energy threshold up to 2~MeV$_{\alpha}$ and one which extends to higher energies induced by $\alpha$ interactions. The first class of event is mostly induced by the external $^{232}$Th sources and it is characterized by:
\begin{equation}
LY_{\beta/\gamma}~=~59.4\pm0.1~keV/MeV.
\end{equation}
About 6~\% of the energy deposited in the crystal is detected as light signal. The fraction of energy converted into light in similar set-ups of other low temperature scintillating bolometers is ca. 8\% for CsI~\cite{CsI} , ca. 2\% for ZnWO$_4$~\cite{ZWO}, ca. 1\% for BGO~\cite{BGO}, demonstrating that this compound has a light output comparable to other well-established scintillating compounds.

We also compute the LY for $\alpha$ interacting particle. In order to evaluate this parameter we take advantage of the internal $\alpha$ contamination of the crystal. The LY$_{\alpha}$ ranges from 8.9$\pm$0.4~keV/MeV for $^{147}$Sm to 11.6$\pm$0.1~keV/MeV for $^{234}$U.

\begin{figure}[h]
\centering
\includegraphics[width=0.5\textwidth]{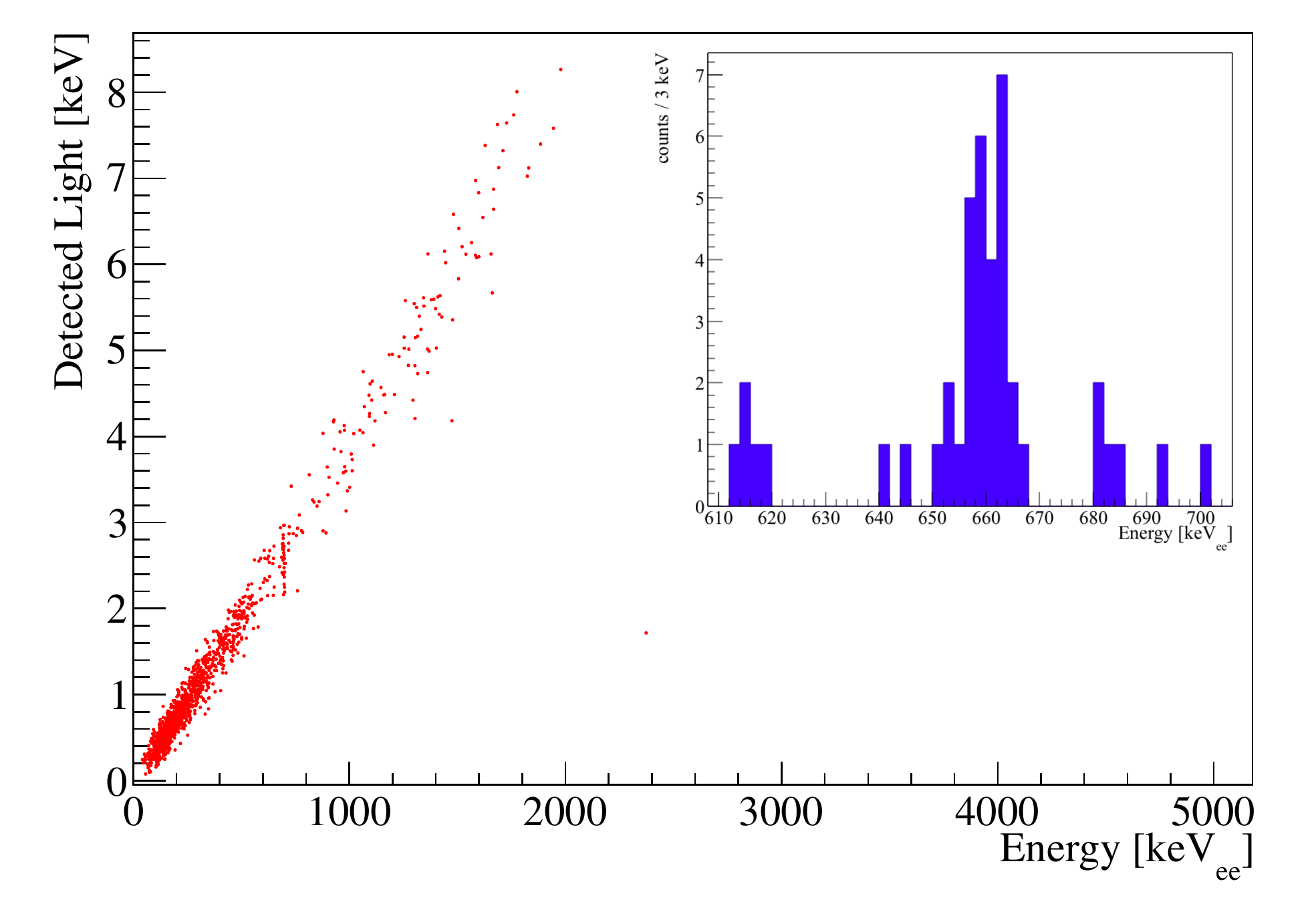}
\caption{Calibration energy scatter plot with a $^{137}$Cs source. In the inset is shown a histogram energy spectrum of $\beta/\gamma$ events produced by the source, in particular the characteristic $\gamma$-line at 661~keV.}
\label{fig:Cs}
\end{figure}

Unfortunately, the small dimensions of the crystal, together with the low intensity of the $^{232}$Th calibration source did not allow us to properly calibrate the $\beta/\gamma$ energy spectrum. 
For this reason we decided to deploy a $^{137}$Cs $\gamma$ source with higher activity and with lower energy to enhance the detection efficiency for its full-energy photoelectric-peak at 661~keV. In Fig.~\ref{fig:Cs}, we show the energy scatter plot for the $^{137}$Cs calibration run, which lasted 2.5~hours. We measure a FWHM energy resolution at 661~keV of 9$\pm$2~keV.

\begin{figure}[t]
\centering
\includegraphics[width=0.5\textwidth]{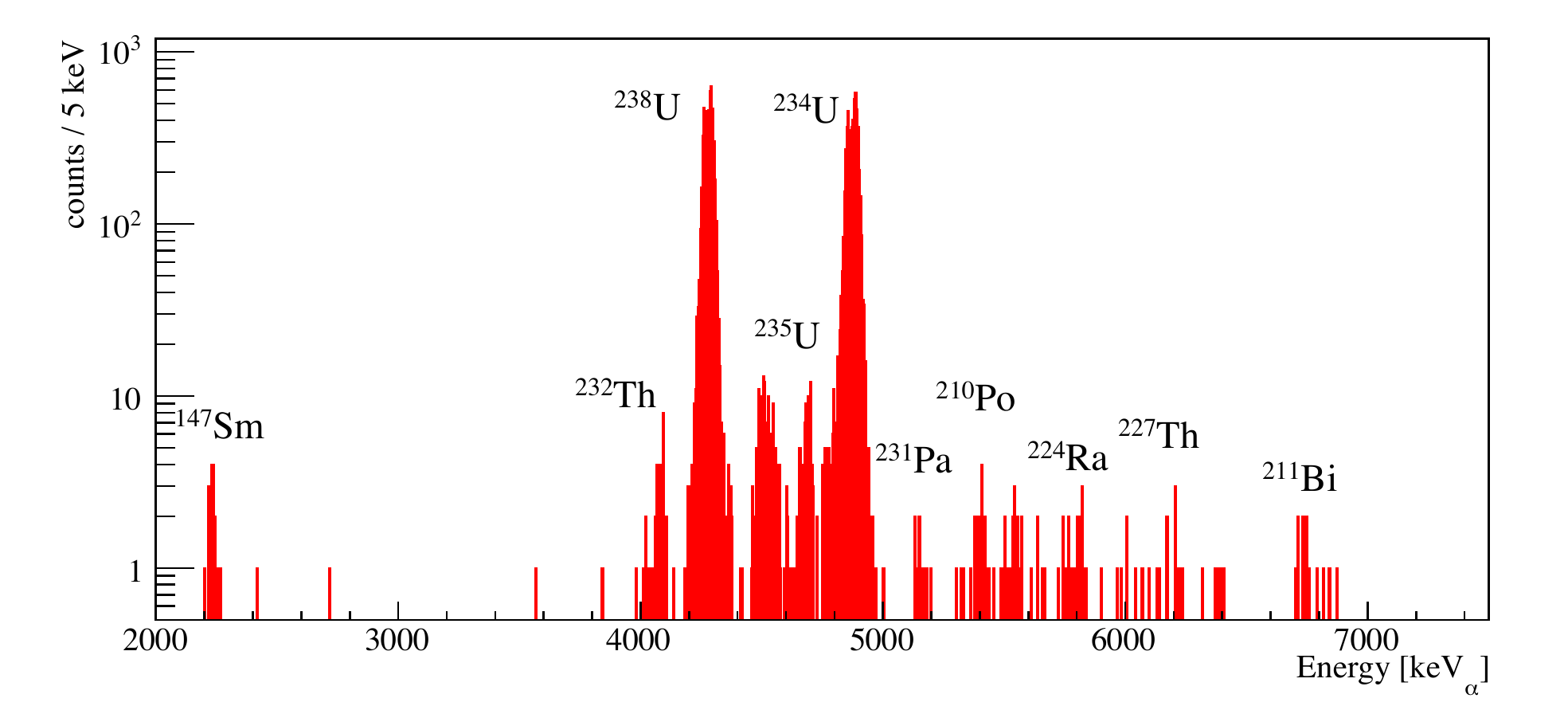}
\caption{YVO$_4$ alpha energy spectrum after applying all cuts for event selection for 115~h of effective data taking. The most relevant peaks are produced by internal radioactive contaminations, i.e. natural radioactivity of $^{232}$Th and $^{238/235}$U.}
\label{fig:alpha}
\end{figure}

The $\alpha$-band in Fig.~\ref{fig:scatter}, in concurrence with Birk's law~\cite{Birks}, has a lower LY compared to the $\beta$/$\gamma$ one. The events in this band are uniquely ascribed to $\alpha$ decays occurring on the detector surface and in the bulk, which produce respectively a continuum of events and some peaks~\cite{sticking}. This energy region is favourable for studying the detector internal radiopurity, because we can benefit from the advantageous signal-to-noise ratio compared to the $\beta$/$\gamma$ region, where the background induced by external $\beta$/$\gamma$ interactions is higher. We select the $\alpha$ band by applying a single cut on the LY$<$20~keV/MeV with an efficiency in rejecting $\beta$/$\gamma$ events $>$~99\%. In Fig.~\ref{fig:alpha} the $\alpha$ energy spectrum of the selected events is shown. The energy scale is calibrated using the known energies of the peaks in the spectrum.\\

In Table~\ref{tab:cont} the internal radioactive contaminations of the crystal are reported. These values are computed by analysing the various $\alpha$ peaks of the energy spectrum shown in Fig.~\ref{fig:alpha}. The high counting rate of the crystal is mainly caused by the rather high U content, both in $^{238}$U and $^{235}$U. If we assume that U is present in the crystal with the natural isotopic abundance~\cite{bohlke2005} ($^{235}$U~:~$^{238}$U = 0.72\%~:~99.27\%) we would expect to measure an activity ratio $^{235}$U/$^{238}$U of 0.046, while the value obtained from the measured activity of the crystal is 0.020. This hints towards a depleted $^{235}$U contamination, which seems to be reasonable according to~\cite{Vanadium}. In fact Vanadium is mostly available on the Earth crust in the form of a carnotite mineral (K$_2$(UO$_2$)$_2$(VO$_4$)$_2$) in Uranium ores. We may speculate that the raw materials used for the crystal production come from U extraction from rocks, and in particular from the enrichment process of $^{235}$U. 

The detector shows also a sizeable $^{147}$Sm contamination of 2.3$\pm$0.5~mBq/kg. This is most probably ascribed to its chemical affinity with Y.

\begin{table}[htp]
\caption{Evaluated internal radioactive contaminations for the YVO$_4$ crystal from $\alpha$ and $\gamma$ spectroscopy, with a bolometric and HP-Ge measurement respectively. Limits evaluated for $^{40}$K, $^{137}$Cs, $^{60}$Co and $^{88}$Y have to be considered as off February 2014. These are evaluated at 90\% C.L.} 
\begin{center}
\begin{tabular}{lccc}
\hline\noalign{\smallskip}
Chain & Nuclide  & Activity & Activity \\ 
            & & [mBq/kg] & [mBq/kg]\\
\noalign{\smallskip}\hline\noalign{\smallskip}
  & & Bolometer & HP-Ge\\
\hline
$^{232}$Th &  \\
 & $^{232}$Th & 8$\pm$1 & -  \\
  & $^{228}$Ra &- & $<$12  \\
   & $^{228}$Th&- &  $<$10  \\
\noalign{\smallskip}\hline\noalign{\smallskip}
$^{238}$U &  \\
 & $^{238}$U & 1373$\pm$2  & -  \\
 & $^{234}$Th&- & 1600$\pm$400  \\ 
 & $^{234m}$Pa&- & 1200$\pm$500  \\
 & $^{234}$U &  1351$\pm$2 & - \\
 & $^{226}$Ra & 5$\pm$1 & $<$12\\
 & $^{210}$Po & 5$\pm$1 & - \\
\noalign{\smallskip}\hline\noalign{\smallskip}
$^{235}$U  & \\
 & $^{235}$U & 56$\pm$5 & 100$\pm$30 \\
\noalign{\smallskip}\hline\noalign{\smallskip}
 & $^{147}$Sm&2.3$\pm$0.5 & -  \\
\noalign{\smallskip}\hline\noalign{\smallskip}
 & $^{40}$K&- & $<$220  \\
\noalign{\smallskip}\hline\noalign{\smallskip}
 & $^{137}$Cs &-& $<$8.6 \\
\noalign{\smallskip}\hline\noalign{\smallskip}
 & $^{60}$Co&- & $<$22 \\
\noalign{\smallskip}\hline\noalign{\smallskip}
& $^{88}$Y&- & $<$7.4   \\
\noalign{\smallskip}\hline\noalign{\smallskip}

\end{tabular}
\label{tab:cont} 
\end{center}
\end{table}

Furthermore a $\gamma$ spectrometric analysis of the YVO$_4$ crystal was performed to evaluate the presence of other radionuclide, which could not be easily investigated with a bolometric measurement, i.e. $\beta$-$\gamma$ cascades. We would like to mention that the evaluation on the concentration of U and Th radionuclides using $\gamma$ spectroscopy was in agreement with the one measured by the bolometric measurement.

The measurement was carried out at the STELLA (SubTerranean Low Level Assay) facility at LNGS. The YVO$_4$ sample was placed directly on the end-cap of a ultra-low-background High-Purity Germanium (HP-Ge) detector. A coaxial p-type Ge detector was used with an active volume of about 450 cm$^3$, and a relative efficiency of 120\%. The detector end-cap is made of a thin Cu window of 1~mm thickness. The detector is surrounded by 25~cm of low-radioactivity lead, 5~cm of copper, and in the inner part of the set-up by few cm of ancient lead. Finally, the shielding are housed in an acrylic box continuously flushed by boil-off N$_2$ gas.
The activity of the investigated radionuclides are reported in Table~\ref{tab:cont}. These were evaluated during a measurement time of 208~h.

\section{Constraints on the development of a bolometric detector for $^{50}$V $\beta^-$ decay investigations}

The expected signals induced by $^{50}$V decay in the YVO$_4$ bolometer are two, as shown in Fig.~\ref{fig:scheme}. The first one is due to the electron capture decay and it is expected to produce a monochromatic line at 1153~keV. On the other side the $\beta^-$ decay, never observed, would induce a continuum of events which extends up to 1038~keV. The decay on the 2$^+$ state will consequently de-excite by a prompt emission, after 8~psec, of a 783~keV $\gamma$ quantum. Most of the $\beta/\gamma$ natural radioactivity ($^{235/238}$U and $^{232}$Th) induce an overwhelming background in the region of interest, as it is the case of the measurement described in this work, thus limiting the sensitivity to the rare decay investigation.

\begin{figure}[h]
\centering
\includegraphics[width=0.4\textwidth]{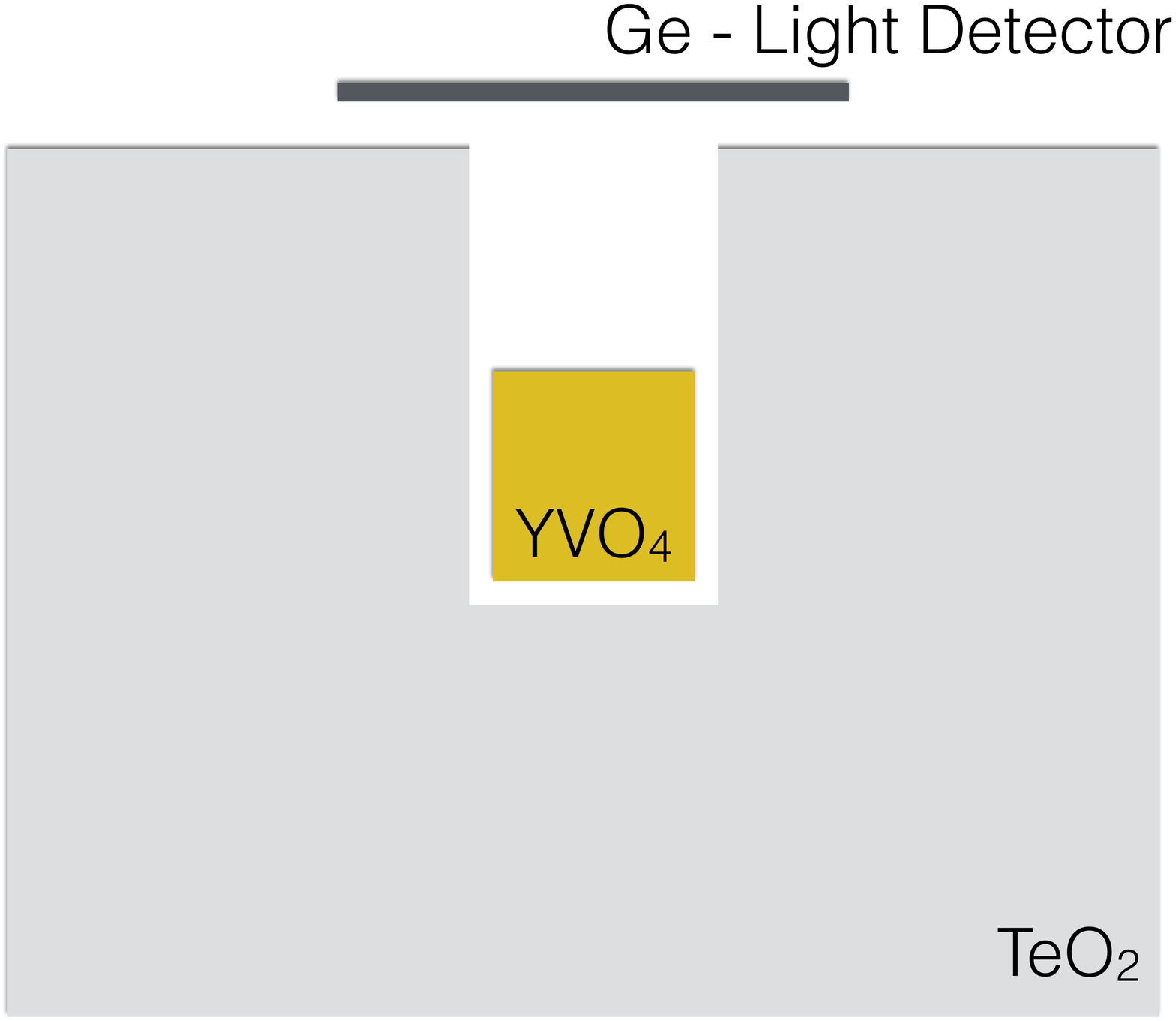}
\caption{Schematic view of a triple-coincidence detector for the positive tagging of $^{50}$V $\beta$/$\gamma$ decay. Any $\beta$ decay in the YVO$_4$ crystal can be identified by means of the heat-light read-out. When the event produced is a $\beta$, then a veto window is open the acquisition of the external veto bolometer which will help in identifying the escaping 783~keV $\gamma$ quantum produced in the $^{50}$Cr de-excitation.}
\label{fig:new}
\end{figure}

In a single YVO$_4$ bolometric detector the $\beta^-$ decay and the de-excitation $\gamma$ will pile-up without inducing a peaking signal. This is caused by the extremely poor time resolution for disentangling the $\beta$-$\gamma$ cascade, for this reason we propose an innovative experimental set-up for the detection of the characteristic 783~keV $\gamma$ quantum, as shown in Fig.~\ref{fig:new}. A YVO$_4$ crystal is operated together with a LD as scintillating bolometer for the identification of $\beta$/$\gamma$ events. Simultaneously a massive high quality crystal, like a TeO$_2$, is operated as bolometric veto. In this configuration the $\beta^-$ decay in the YVO$_4$ acts as a trigger for the TeO$_2$ detector that has to absorb the 783~keV $\gamma$ quantum. We take advantage of the different mean-free path of the $\beta$ and $\gamma$ in the detector source. TeO$_2$ crystal is the best choice as external veto detector given its high radiopurity~\cite{TeO2_2vDBD}, excellent energy resolution~\cite{CUORE-0_detector} and availability of massive size~\cite{TeO2_massive}.

We ran a MonteCarlo simulation where the experimental set-up is the same shown in Fig.~\ref{fig:new}, where the YVO$_4$ crystal has the same dimensions as the crystal described in this work and the TeO$_2$ crystal is a cube with side of 10~cm. We generate the $\beta/\gamma$ cascade in the YVO$_4$ crystal and we evaluate the detection efficiency of the expected signals. About 90\% of the simulated $\beta$ events will be confined in the YVO$_4$ crystal without inducing a signal in the TeO$_2$ veto, while the detection efficiency of the 783~keV $\gamma$ in the TeO$_2$ is about 40\%. These values hint towards a feasibility of the detection of the $^{50}$V $\beta^-$ decay observation with this innovative experimental set-up.

Nevertheless a thorough assay of radiopure raw materials for the production of high purity YVO$_4$ crystals has to be carried out in order to perform a highly sensitive investigation of the 4-fold forbidden decay. In fact, crystals available on the market, as the one operated in this work do not meet the requirements for such studies. The high internal $^{235/238}$U and $^{232}$Th concentration underline the need for the development of purification techniques of raw materials for the growth of radiopure YVO$_4$ crystals. Innovative purification procedures are currently under development~\cite{serge}. One of the most promising technique is electron-beam melting that is able to reduce by a factor 10 and 100, respectively for U and Th contaminations, in a Vanadium metal sample.

\section{Conclusions}
We operated for the first time a YVO$_4$ as a cryogenic scintillating bolometer, it showed great bolometric performance and an excellent light output. On the other hand the radiopurity of the detector is compromised by a high $^{235/238}$U concentration which currently limits its sensitivity for the the investigation of the $^{50}$V $\beta^-$ decay branch. Currently cutting-edge technologies developed for the purification of raw materials for the production of radiopure YVO$_4$ crystals, which may help in suppressing the background induced by natural radioactivity. We propose an innovative approach for an efficient detection of the characteristic de-excitation $\gamma$ following $^{50}$V $\beta^-$ decay using a triple-coincidence system. Preliminary MonteCarlo simulations demonstrate the feasibility of such a technique.


\begin{thebibliography}{99}
%
%

\bibitem{Bi209}
P.~de~Marcillac et al., Nature {\bf422}, (2003) 876.

\bibitem{Bi209exc}
J.W.~Beeman et al., Phys. Rev. Lett. {\bf108}, (2012) 062501.

\bibitem{Iachello}
J.~Barea et al., Phys. Rev. C {\bf87}, (2013) 014315.

\bibitem{Suhonen}
J.~Kostensalo et al., Phys. Rev. C {\bf95}, (2017) 044313.

\bibitem{1V}
J.~Heintze, Z. Naturforschung A {\bf10}, (1955) 77

\bibitem{2V}
R.N.~Glover et al., Philos. Mag. {\bf2}, (1957) 697.

\bibitem{3V}
E.R.~Bauminger et al., Phys. Rev. {\bf110}, (1958) 953.
 
\bibitem{4V}
A.~McNair, Philos. Mag. {\bf6}, (1961) 559.

\bibitem{5V}
D.E.~Watt et al., Nucl. Phys. {\bf29}, (1962) 648.
 	
\bibitem{6V}
C.~Sonntag et al., Z. Phys. {\bf197}, (1966) 300.
 	
\bibitem{7V}
A.~Pape et al., Phys. Rev. C {\bf15}, (1977) 1937.
 	
\bibitem{8V}
D.E.~Alburger et al., Phys. Rev. C {\bf29} (1984) 2294.
 	
\bibitem{9V}
J.J.~Simpson et al., Phys. Rev. C {\bf31} (1985) 575.
 	
\bibitem{10V}
J.J.~Simpson et al., Phys. Rev. C {\bf31}, (1989) 2367.
 	
\bibitem{11V}
H.~Dombrowski et al., Phys. Rev. C {\bf83}, (2011) 054322.

\bibitem{Pirro_rev}
S.~Pirro et al, Annu. Rev. Nucl. Part. Sci. {\bf67} (2017) 161-181.

\bibitem{V50}
G.~Audi et al., Nucl. Phys. A {\bf729}, (2003) 337.

\bibitem{W180}
C.~Cozzini et al., Phys. Rev. C {\bf70}, (2004) 064606.

\bibitem{pbwo4}
J.W.~Beeman et al., Eur. Phys. J. A {\bf49}, (2013) 50.

\bibitem{Eu151}
N.~Casali et al., J. Phys. G {\bf41}, 075101 (2014).

\bibitem{Sm148}
N.~Casali et al. J. Low Temp. Phys. {\bf184}, (2016) 952.

\bibitem{LiMoNe}
L.~Cardani et al., J. Instrum. {\bf8}, (2013) P10002.

\bibitem{LD_characterization}
J.W.~Beeman et al., J. Instrum. 8 (2013) P07021.

\bibitem{muon}
E.~Andreotti et al., Astropart. Phys. {\bf34}, (2010) 18-24.

\bibitem{Zn82Se}
D.R.~Artusa et al., Eur. Phys. J. C {\bf76}, (2016) 364.

\bibitem{light_sync}
G.~Piperno et al., J. Instrum. {\bf6}, (2011) P10005.

\bibitem{OF}
E.~Gatti et al., Rivista del Nuovo Cimento {\bf9}, (1986) 1.

\bibitem{CsI}
G.~Angloher et al., Astropart. Phys. {\bf84}, (2016) 70-77.

\bibitem{ZWO}
M.~Kiefer et al., Nucl. Instrum. Meth. A {\bf821}, (2016) 116-121.

\bibitem{BGO}
J.W.~Beeman et al., Phys. Rev. Lett. {\bf108}, (2012) 062501.

\bibitem{Birks}
J.B.~Birks, Proc. Phys. Soc. A {\bf64}, (1951) 874.

\bibitem{sticking}
M.~Clemenza et al., Eur. Phys. J. C {\bf71}, (2011) 1805.

\bibitem{bohlke2005}
M.~Berglund et al., Pure Appl. Chem., {\bf83}, (2011) 397.

\bibitem{Vanadium}
J.G.H.~du~Preez, Radiat. Prot. Dosim. {\bf26}, (1989) 7-13.

\bibitem{EC_V50}
H.~Dombrovski et al., Phys. Rev. C {\bf83}, (2011) 054322.

\bibitem{TeO2_2vDBD}
C.~Alduino et al,  Eur. Phys. J. C {\bf77}, (2017) 13.

\bibitem{CUORE-0_detector}
C.~Alduino et al., J. Instrum. {\bf7}, (2016) P07009.

\bibitem{TeO2_massive}
L.~Cardani et al., J. Instrum. {\bf7}, (2012) P01020.

\bibitem{serge}
Yu.P.~Bobrov et al., Voprosy Atomnoj Nauki i Tekhniki  {\bf46}, (2014) 27-31.

\end{thebibliography}
\end{document}